\documentclass[12pt,a4paper,twoside]{amsart}
\usepackage{amsmath,bbm,amssymb,amsxtra}
\usepackage{enumerate}
\allowdisplaybreaks
\usepackage{epsfig}
\usepackage{ae}
\theoremstyle{definition}

\newtheorem{lemma}{Lemma}[section]
\newtheorem{proposition}[lemma]{Proposition}

\newtheorem{theorem}[lemma]{Theorem}

\newtheorem{definition}[lemma]{Definition}

\newcommand{\prop}[1]{\begin{proposition}\label{#1}
\sl }
\newcommand{\eprop}{\end{proposition}}
\newcommand{\thm}[1]{\begin{theorem}\label{#1}
\sl }
\newcommand{\ethm}{\end{theorem}}
\newcommand{\lem}[1]{\begin{lemma}\label{#1}
\sl }
\newcommand{\elem}{\end{lemma}}

\newcommand{\defin}[1]{\begin{definition}\label{#1}
\sl }
\newcommand{\edefin}{\end{definition}}

\newcommand{\beqno}{\begin{eqnarray*}}
\newcommand{\eeqno}{\end{eqnarray*}}
\newcommand{\beqla}[1] {\begin {eqnarray}\label{#1}}
\def\eeq {\end {eqnarray}}
\newcommand{\beq}{\begin {eqnarray}}

\newcommand{\E}{\mathbb{E}}

\newcommand{\R}{\mathbb{R}}

\newcommand{\IZ}{\mathbb{Z}}

\newcommand{\supp}{{{\rm supp}\,}}

\def\CA{{\mathcal A}}              
\def\CD{{\mathcal D}}       \def\CE{{\mathcal E}}       \def\CF{{\mathcal F}}
       \def\CH{{\mathcal H}}       
       \def\CK{{\mathcal K}}       \def\CL{{\mathcal L}}
              
\def\CP{{\mathcal P}}

\def\qq{ \begin{eqnarray} }
\def\qqq{ \end{eqnarray} }
\def\rr{ \begin{equation} }
\def\rrr{ \end{equation} }
\def\non{ \nonumber }
\def\qq{ \begin{eqnarray} }
\def\qqq{ \end{eqnarray} }
\def\non{ \nonumber }

\newcommand{\tr}{\hbox{tr}}

\newcommand{\hf}{{_1\over^2}}
\newcommand{\BbbT}{\mathbb{T}}

\begin{document}
\title[ERGODICITY OF TWO DIMENSIONAL TURBULENCE]
{ERGODICITY OF TWO DIMENSIONAL TURBULENCE}
\author{Antti KUPIAINEN}
\address{Department of Mathematics\\
University of Helsinki\\
P.O. Box 68 (Gustaf H\"allstr\"omin katu 2b)\\
FI-00014 Helsinki -- Finland}
\email{ajkupiai@mappi.helsinki.fi}

\maketitle

%%%%%%%%%%%%%%%%%%%%%%%%%%%%%%%%%%%%%%%%%%%%%%%%%%%%%%%%%%%%%%%%%%%%%%%%%%%%%%%%%
\begin{abstract}
We review the work of Hairer and Mattingly on ergodicity of two
dimensional Navier-Stokes dynamics and discuss some open mathematical
problems in the theory of 2d turbulence.
\end{abstract}

\section{Introduction}
\label{se:intro}

The problem of turbulence has been described as the last great unsolved 
problem of classical physics. Understanding 
of the complicated motion
of fluids in the presence of obstacles or stirring has been a challenge
to mathematicians, physicists and engineers for quite a time now.
The equations governing macroscopic fluid motion, the Navier Stokes equations, have
been known for close to two centuries.
For an incompressible fluid 
in units where the density equals one
they read
\qq
\partial_t u +u\cdot\nabla u
=
\nu\Delta u-\nabla p+f.
\label{ns}
\qqq 
$u(t,x)\in\R^d$ is the velocity field at time $t$ at $x\in\Lambda$, a domain in $\R^d$
subject to the incompressibility condition
\qq
\nabla\cdot u=0 
\label{inco}
\qqq 
and suitable boundary conditions on $\partial\Lambda$. $\nu$ is the viscosity coefficient
of the fluid, $p(t,x)$  the pressure
and $f(t,x)$ the external force that sustains the flow. Given $f$ and $u(0,\cdot)$ the task is
to find $u$ and $p$.
It is fair to say that theoretical understanding of the  
consequences of these equations is still in its infancy. On the mathematical side,
existence of smooth solutions for the three dimensional
NS equations is wide
open and has been chosen by some  as one of the major problems of
mathematics \cite{Clay}. On the physical side, experimental
violations of the Kolmogorov scaling theory of turbulence \cite{Fri} are
still waiting for theoretical understanding.

In two dimensions, i.e. for flows on the plane, there has been some progress
during the last ten years. On the physical side, 2d turbulence has been the subject
of accurate numerical and experimental studies \cite{Bof}, \cite{RDCE} and mathematically
the ergodic theory of the NS flow has been under intensive study.  

It
is important to realize that for the problem of turbulence 
one is interested in a very particular kind of force in (\ref{ns}), namely
one that has a fixed { length scale $L$} built into it. Examples
of this are flows past obstacles, with $L$  the characteristic
size of the obstacle. In such a setup the flow exhibits universal statistical
properties as the viscosity parameter tends to zero (actually the
control parameter is a
dimensionless quantity, the Reynolds number given by $\frac{Lv}{\nu}$ where $v$
is a velocity scale related to the forcing).
E.g. time averages of measurements of suitable functions
of $u$ seem to show statistical properties only depending on
the Reynolds number. It is therefore of some interest to inquire
about the foundations for such statistical studies i.e. about the
ergodic properties of the NS flow in the turbulent setup of
a fixed scale high Reynolds number forcing.

A convenient model for isotropic and homogenous turbulence
(i.e.
in the limit of large Reynolds number and away from the boundary $\partial\Lambda$)
is to consider \footnote{To get to the turbulent state
one actually has to modify  (\ref{ns}) a bit, see Section 8} eq. (\ref{ns}) on the torus $\BbbT^2=\R^2/(2\pi\IZ)^2$ and take $f$
random, a Fourier series with finite number of terms and coefficients
independent white noises (see below).  Then the deterministic dynamics
of  (\ref{ns}) is replaced by a Markov process and one may pose 
questions on its ergodic properties: whether the process has a unique stationary
state and whether this is reached and with what rate from arbitrary
initial conditions.

This Markov process is a diffusion process of a very degenerate type.
While the phase space is infinite dimensional the noise is finite dimensional.
There are two general mechanisms that can contribute to the ergodic and
mixing properties of stochastic flows. One is dissipation, coming in our
case from the Laplacean in   (\ref{ns}). Dissipation contributes to ergodicity by
exponential contraction of phase space under the flow. Second mechanism
comes from the spreading of the noise from its finite dimensional subspace 
due to the nonlinear term in  (\ref{ns}). In finite dimensional diffusion processes
this leads to hypoellipticity if the noise spreads to the full phase space: the
transition kernels are smooth (for equations with smooth coefficients). 
  Combined
with some irreducibility of the process ergodicity follows. 

In our infinite dimensional setup the dissipation due to the Laplacean
leads to strong damping of large enough (depending on the Reynolds number) Fourier modes.
If we keep noise on all the other, low, modes then one can reduce the
problem to a low mode dynamics, albeit with some (exponentially decaying) memory
due to the large modes. Proofs of ergodicity and mixing of the dynamics
were given  in this case in the works \cite{BKL}, \cite{EMS} and \cite{KS}.
However,  it seemed far from trivial to extend the hypoellipticity ideas to
the infinite dimensional setup to control also the case of
very degenerate forcing where the number of
forced modes does not depend on the Reynolds number. This was accomplished by Hairer and
Mattingly \cite{HM1},   \cite{HM2} who gave sharp sufficient conditions
for the noise to produce ergodic and mixing dynamics. In what follows I will present
the main points of their approach focusing on the difference to  finite dimensional
hypoelliptic diffusions. The papers  \cite{HM1},   \cite{HM2} are very clearly
written and they contain plenty of background material, especially \cite{HM2} 
which also builds a more general formalism applicable also to some
reaction-diffusion equations. \cite{HM2} also corrects a mistake in
 \cite{HM1} so it should  be
consulted for a thorough study.
In the final section I discuss more informally what we have learned about
2d turbulence and what issues might be accessible to a rigorous  mathematical
analysis.

I would like to thank J. Bricmont, M. Hairer and J. Mattingly for comments on this exposition
and the European Research Council and Academy of Finland for financial support.

\section{2D NS EQUATIONS}

The fundamental fact that is behind both the mathematical and physical
understanding of 3d NS equations is energy conservation: in the absence
of forces smooth inviscid flow preserves the $L^2$ norm of $u(t,\cdot)$.
In two dimensions there is a second conserved quantity, the  {\it enstrophy},
which is related to the $H^1$
norm and which leads to quite different physics and to much better regularity.

Let us first
define the {\it vorticity}  
$$
\omega = \nabla\times u,
$$
which in $d=2$ is a (pseudo)scalar: $\omega =
\partial_1u_2 -\partial_2u_1$. The NS equation becomes in terms of $\omega$ a transport equation:
\qq
\dot \omega = \nu \Delta \omega - u\cdot \nabla \omega + g.
\label{o}
\qqq
where $g = \partial_1 f_2 - \partial_2 f_1$. We will assume
the average force vanishes i.e. $\int f(t,x)dx=0$. Then (\ref{ns})
preserves the condition $\int u(t,x)dx=0$ which we will assume.
  The incompressibility condition (\ref{inco}) allows to
  write $u=\CA\omega$ where the linear operator
  $\CA$ is given in terms of the Fourier transform by
  \qq
\widehat {\CA\omega}(k)=i(k_2,-k_1)k^{-2}\hat\omega(k)
\label{L}
\qqq
for $k\in\IZ^2\setminus 0$.

The {enstrophy} $\CE$
is defined to be (half of) the $L^2$-norm of $\omega$:
$$
\CE = \hf\int \omega
(t,x)^2dx:=\hf\|\omega(t)\|^2$$
 For a smooth $u$ the condition
 $\nabla\cdot u=0$  leads to the absence of contribution
from the nonlinear term to the evolution of the enstrophy:
\qq
{d\CE\over dt} = -\nu \int (\nabla\omega)^2dx
 + \int \omega gdx.
\label{ens}
\qqq 
where the first term on the RHS can be interpreted as an
enstrophy  dissipation  rate
and the second one as an
enstrophy injection rate.  Using Poincare inequality $\|\nabla\omega\|\geq \|\omega\|$ 
and simple estimates one deduces
\qq
\Vert \omega (t)\Vert^2\leq e^{-\nu t}\Vert \omega (0)\Vert^2
+  \nu^{-2}\sup\limits_t \Vert g(t)\Vert^2.
\label{poinc}
\qqq
This  a'priori estimate for the
$H^1$
norm of $u$ is the main ingredient in the proof of global regularity of the
2d NS flow.

We wish now to discuss a version of ({\ref{o}) where the force $g$ is random.
We work in the subspace  of real valued $L^2(\BbbT^2)$ functions with $\hat\omega(0)=0$.
It will be convenient to use the following basis for this space.
Let $ Z^+$ be the "upper half plane" in $\IZ^2$
consisting of $k=(k_1,k_2)$ with $k_2>0$ or $k_2=0$ and $k_1>0$. Hence
 $\IZ^2\setminus 0=Z^+\cup (-Z^+)$. Let $e_k=\sin kx$ for $k\in Z^+$
 and $e_k=\cos kx$ for  $-k\in Z^+$. 
For each $k\in \IZ^2$ pick independent 
Brownian motions  $\beta_k(t)$ with unit speed, denoted collectively by $\beta(t)$ and numbers
$\gamma_k\in\R$.  Let
\qq
 Q\beta(t)=\sum_{k\in \IZ^2} \gamma_k\beta_k(t)e_k.
\label{B}
\qqq 
The stochastic version of  eq. ({\ref{o}) reads
\qq
d\omega = (\nu \Delta \omega - u\cdot \nabla \omega )dt+ Qd\beta.
\label{o1}
\qqq
Regularity of the stochastic flow proceeds in parallel with the deterministic case
as long as $\gamma_k$ have enough decay at infinity. The analog of the enstrophy
conservation eq. (\ref{ens}) is obtained by an application of the Ito formula
\qq
d\CE =\hf d\|\omega\|^2= -\nu \|\nabla\omega\|^2dt+(\omega,Qd\beta)
 + \epsilon dt
\label{ito}
\qqq 
where $\epsilon=2\pi^2\sum_k\gamma_k^2$ can be interpreted as the enstrophy injection rate.
Taking averages we get a probabilistic analog of   (\ref{ens}) and (\ref{poinc}):
\qq
{d\over dt}\E\CE = -\nu\E  \|\nabla\omega\|^2
 + \epsilon.
\label{enspro}
\qqq 
and
\qq
\E\Vert \omega (t)\Vert^2 \leq e^{-2\nu t}\Vert \omega (0)\Vert^2_2
+  \nu^{-1}\epsilon.
\label{ens10}
\qqq
%\qq{d\over dt}\E\ \CE = -\nu  \|\nabla\omega\|_2^2 + \epsilon\label{ens}\qqq 
  Actually (\ref{ito}) can be used to control exponential 
moments of the enstrophy \cite{BKL}, \cite{HM1} Lemma A.1.:
\qq
\E\exp(\eta\|\omega(t)\|^2 )\leq
2\exp(\eta e^{-\nu t}\|\omega(0)\|^2)
\label{ensexp}
\qqq
for all $\eta\leq \frac{\nu}{\epsilon}$
i.e. probability for large $L^2$ norm is exponentially small. 
(\ref{ito}) also allows to control the time integral of the $H^1$ norm:
\qq
\E\exp(\eta\nu\int_0^t\|\nabla\omega(t)\|^2 )\leq
2\exp(\eta\epsilon t+\eta\|\omega(0)\|^2)
\label{ens1}
\qqq
again for all $\eta\leq \frac{\nu}{\epsilon}$.
Such
a'priori estimates allow one to prove the existence and pathwise
uniqueness of strong solutions to eq. (\ref{o1}) under quite
general conditions on the noise coefficients $\gamma_k$, see e.g.
\cite{Fla} and \cite{MR}. Of course the PDE regularity is the harder
the less decay the $\gamma_k$ have at infinity. As explained
in the introduction, for the turbulence problem only finite number
of $\gamma_k$ are nonzero. Thus from the  point of view of
regularity the turbulent case is easy (this is not true in 3d!).
However, the ergodicity of the flow becomes the harder the less there is noise.

\section{INVARIANT MEASURE}

Let us now specialize to the case where $\gamma_k=0$ for $k\notin K$
where $K$ is a finite set. Thus the noise is finite dimensional: $\beta =\{\beta_k\}_{k\in
K}$ can be identified with the Wiener process
in $\Omega=C([0,\infty), \R^D)$ where $D= {|K|}$, equipped with the Wiener measure $W(db)$.
The solution of eq. (\ref{o1}) is a
one parameter family of continuous maps $\Phi_t:\Omega\times L^2(\BbbT^2)
\to L^2(\BbbT^2)$ such that $\omega(t)=\Phi_t(\beta,\omega_0)$ solves
 eq. (\ref{o1})  with initial condition $\omega_0$ and noise realization $\beta$. Actually, $\Phi_t$
 is (Fr\'echet) differentiable in $\beta$ and $\omega_0$.
 
 $\omega(t)$ is a Markov process with state space $ H=L^2(\BbbT^2)$.
It gives rise to transition
probabilities $P_t(\omega_0,A)$ which are probability measures on
$H$, giving the probability of entering the set $A\subset H$ at
time $t$ given that at time $0$  we have $\omega(0)=\omega_0$:
$$P_t(\omega_0,A) = \E 1_A(\omega(t)).$$
The transition probabilities generate a semigroup $\CP_t$ on bounded
measurable functions on $H$ by the same formula:
\qq
 P_t\phi=\int P_t(\cdot,d\omega)\phi(\omega)
\label{semi}
\qqq 
and the adjoint semigroup acting on bounded (Borel) measures:
\qq
 P_t^*\mu=\int\mu (d\omega_0)P_t(\omega_0,\cdot)\label{semi*}
\qqq 
We are interested in the invariant (or stationary) probability measures $\mu^*$ satisfying the
equation
\qq
P_t^*\mu^* = \mu^*.
\label{stati}
\qqq
Existence of an invariant measure is straightforward given the
strong probabilistic control of the flow. One considers the family of
time averages $\mu^{(\omega_0)}_t=t^{-1}\int_0^tdsP_s(\omega_0,\cdot)$ and
shows it is tight. Prohorov's theorem then yields a limit point
which is shown to be invariant.

Uniqueness of the invariant measure is much more subtle.
It implies {\it ergodicity}, i.e. in particular the equivalence of time
averages and ensemble averages: $ \lim_{t\to\infty} \mu^{(\omega_0)}_t(\phi)
=\mu(\phi)$ for all $\phi\in L^2(H,\mu)$ and $\mu$-a.s. in $\omega_0$. 
In practice one would like to have more i.e. the convergence in some sense
of the measures $P_t(\omega_0,\cdot)$ to $\mu^*$ as $t\to\infty$. This
leads to various {\it mixing} concepts.

\section{DISSIPATION AND SMOOTHING}

For finite dimensional diffusion processes it is well known that
uniqueness of the invariant measure follows from recurrence
and smoothing properties of the transition probabilities. Let us
sketch a special version of this argument having the application to 
NS in mind.

The semigroup $\CP_t$ is called {\it strong Feller} if the image is continuous
for $\phi$ measurable. This has drastic consequences for the supports
of invariant measures. Recall that $x$ belongs to the support of
a finite Borel measure $\mu$ on a Polish space (our setup) if
$\mu(U)>0$ for all open $U$ containing $x$. Then {\it the supports
of two distinct ergodic invariant probability measures for a strong Feller semigroup
are disjoint}. To see this, suppose $\mu\perp\nu$ and  $x\in \supp \mu\cap
\supp \nu$. Pick $A$ with $\mu(A)=1$ and $\nu(A)=0$.  By strong Feller
there exists a $U$ containing $x$ such that $\sup_{y,z\in U}|P_t(y,A)-
P_t(z,A)|\leq\hf$. Moreover by assumption $\alpha:=\min\{\mu(U),\nu(U)\}>0$.
Write $\mu=(1-\alpha)\bar\mu+\alpha\mu_U$ with $\bar\mu$ and $\mu_U$
probability measures with $\mu_U(U)=1$ (i.e.$\mu_U=\mu 1_U/\mu(U)$)
and $\nu$ similarly. Then, by invariance $|\mu(A)-\nu(A)|=|\CP_t^*\mu(A)-\CP_t^*\nu(A)|$
and thus
\qq
 1&=&|\mu(A)-\nu(A)|\leq (1-\alpha)
 |\CP_t^*\bar\mu(A)-\CP_t^*\bar\nu(A)|+\alpha|\CP_t^*\mu_U(A)-\CP_t^*\nu_U(A)|\non\\
 &\leq&(1-\alpha)+\alpha\int_{U\times U}|\CP_t(y,A)-\CP_t(z,A)|\mu_U(dy)\nu_U(dz)
 \leq 1-\hf\alpha
\non
\qqq 
a contradiction.

Suppose now that we knew that there exists an $x$ that necessarily
belongs to the support of every invariant measure of a strong Feller
semigroup. We could then conclude uniqueness. This is a reasonable
strategy for the NS equation. Indeed, $\omega=0$ is such a point. This
follows since NS equation is dissipative. Without forcing 
the fluid slows down i.e. the $L^2$ norm decays exponentially
(see eq. (\ref{poinc})).  There
is a non zero probability for the force to stay small enough so that
any neighborhood of $0$ can be reached.
 More precisely, the Ito formula (\ref{enspro}) combined
with Poincare inequality $\|\omega\|\leq\|\nabla\omega\|$ yields
$$
\int\mu(d\omega)\|\omega\|^2\leq \epsilon/\nu
$$
for every invariant probability measure $\mu$. Hence, there exists $R<\infty$
such that every such measure has at least half its mass in the
ball $B_R$ of radius $R$ centered at $0$ in $H$. Thus one needs to show: for all $r>0$
there exits $T_r<\infty$ such that
$$
I_r:=\inf_{\omega_0\in B_R}P_{T_r}(\omega_0, B_r)>0
$$
(see \cite{EM}, Lemma 3.1.). Then $\mu(B_r)=\CP_t^*\mu(B_r)\geq\hf I_r>0$ for all $r>0$.

This strategy does not quite work in our case since the
strong Feller property is very hard to show for  $\CP_t$ and
might very well not be true. One of the main accomplishments
of Hairer and Mattingly was to replace it with a condition
that is more natural for NS and yet allows one to conclude
 that the supports of invariant measures are disjoint.

\section{ASYMPTOTIC STRONG FELLER PROPERTY}

A strong Feller semigroup maps bounded functions to
continuous ones. Often the easiest
way to prove this is to show a bit more \cite{DPZ}, Lemma 7.1.5.:
%\begin{prop} 
\prop{4.1.}
A semigroup on a Hilbert space H is strong Feller if 
for all
$\phi:H \rightarrow \R$ with $\|\phi\|_\infty :=\sup_{x \in H}|\phi(x)|$ 
and $\|D \phi\|_\infty$ finite
one has
\begin{equation}
        \|D \CP_{t}\phi(x)\| \leq C(\|x\|) \|\phi\|_\infty\;,
        \label{sf}
\end{equation}
where $C:\R_+ \rightarrow \R$ and $D$ is the Frechet derivative.
 \eprop% \end{prop}
We will now argue that the condition (\ref{sf}) is not very natural for
the NS dynamics.
As mentioned in the introduction there are (at least) two ways ergodicity
can result. One is due to smoothing by the noise, the other is due to
dissipation that erases memory of the initial conditions. The former effect 
leads to a condition like (\ref{sf}), the latter not. Let us next
discuss the latter effect in our case. 

Let $J_{s,t}$ with $s< t$ be the
derivative of the NS flow (\ref{o1})  between times $s$ and $t$, i.e.\  for every $\xi \in H$,
$J_{s,t}\xi:= \xi(t)$ is the solution of the linear equation
\begin{equation}
\partial_t \xi(t) = \nu \Delta \xi(t)+ \CA\omega(t)\cdot\nabla\xi(t)+
\CA\xi(t)\cdot\nabla\omega(t):=\CL_{\omega(t)}\xi(t)
\label{JLinear}
\end{equation}
for $t>s$ and $ \xi(s) = \xi$.
This linear equation is readily controlled in terms of the $H^1$ norm of
$\omega$ (\cite{HM1}, Lemma 4.10):
\begin{equation}
\|\xi(t)\|\leq \exp\big({C(\delta,\nu)(t-s)+\delta \int_s^t\|\nabla\omega(r)\|^2dr}\big)\|\xi(s)\|
\label{JLinear1}
\end{equation}
for any $\delta>0$. Combining with the a priori estimate (\ref{ens1}) then
\begin{equation}\label{JLinear2}
\E \|J_{s,t}\|^p\leq 2^p\exp\big(C(\epsilon,\nu,\eta,p)(t-s)+\eta\|\omega(s)\|^2\big)
\end{equation}
for all $\eta>0$, all $p<\infty$.

Eqs. (\ref{JLinear1}) and (\ref{JLinear2}) indicate possible exponential separation of
trajectories. However, since the Laplacean is the Fourier multiplier $-k^2$ 
 it is not surprising that
the high Fourier modes of $\xi$ are strongly damped for a time
that can be taken as large as we wish as $N$ is increased. This is expressed by \cite{HM1}, Lemma 4.17.:
%\begin{lemm} \label{high}
\lem{le:4.2}
  For every $p\geq 1${\rm ,} every $T>0${\rm ,} and every two constants $\gamma,
  \eta > 0${\rm ,} there exists an orthogonal projector $\pi_\ell$ onto a finite number of
  Fourier modes such that
\qq  \E( \|(1-\pi_\ell^{}) J_{0,T}^{}\|^{p}+
  \|J_{0,T}^{}(1-\pi_\ell^{})\|^{p})  \le \gamma
  e^{\eta \|w_0\|^2}\non
  \qqq
  \elem
%\end{lemm}
For such contracting dynamics (\ref{sf}) is not a natural condition
to try to prove. Indeed, let $\xi_h=(1-\pi_\ell)\xi$ be the
projection of $\xi$ to the high modes and 
consider the toy problem where we apply $(1-\pi_\ell)$ to eq. (\ref{JLinear}) 
and drop altogether the $\omega$-dependent terms: 
$$\partial_t \xi_h(t) = \nu \Delta \xi_h(t).$$
Then for a function
$\phi(\omega)=\psi((1- \pi_\ell)\omega)$ depending only on the high modes
we have $D\CP_{t}\phi(\omega_0)\xi=\E D\phi(\omega(t))\xi_h(t)$.
Since in this toy case $\|\xi_h(t)\|\leq e^{-At}\|\xi\|$ for $A>0$ we conclude
\qq
  \|D \CP_{t}\phi(x)\| \leq e^{-At} \|D\phi\|_\infty.
%\label{}
\non
\qqq 
This toy model and Lemma 5.2.  motivate the following definition by Hairer and Mattingly 
( \cite{HM1},  Proposition 3.12.)
%\begin{defi} 
\defin
A semigroup $\CP_t$ on a Hilbert space $\CH$ is {\bf asymptotically strong Feller} if{\rm ,} 
there exist two positive sequences $t_n$ and $\delta_n$   with
$\{t_n\}$ nondecrea\-sing and $\{\delta_n\}$ converging to zero such that for all
$\phi:\CH\rightarrow \R$ with $\|\phi\|_\infty$ and $\|D\phi\|_\infty$ finite{\rm ,}
\begin{equation} \label{SF}
|D \CP_{t_n}\phi(x)| \leq C(\|x\|) \bigl( \|\phi\|_\infty + \delta_n \|D \phi\|_\infty\bigr) 
\end{equation}   
for all $n${\rm ,} where $C:\R_+ \rightarrow \R$.
%\end{defi}
\edefin
(Hairer and Mattingly actually give a "topological" definition of asymptotically strong Feller
condition which is implied by the one above). The main point is the following
result whose proof is similar to the one given above  in the strong Feller case ( \cite{HM1}, Theorem 3.16):
\prop If the
semigroup is asymptotically strong Feller at $x$ then $x$ belongs to the
support of at most one ergodic invariant measure .
\eprop

We saw above that it is not unreasonable to expect  that the high mode dynamics
give rise to the second term in (\ref{SF}). Thus the question remains: why would
the low mode dynamics be strong Feller? The answer to this question lies
in the {\it hypoellipticity} of the low mode dynamics.

\section{HYPOELLIPTICITY}

Let us think about the low mode dynamics first in the {\it 
Galerkin approximation} i.e. by putting the high modes to
zero. More formally, consider the equation
\qq
d\omega =( \nu \Delta \omega - \pi_\ell(u\cdot \nabla \omega ))dt+ Qd\beta.
\label{gal}
\qqq
where we assume the forcing is on low modes $(1-\pi_\ell) Q\beta=0$ and
set $(1-\pi_\ell) \omega=0$. Eq. (\ref{gal}) defines a diffusion process
in a finite dimensional space which we may identify with $\R^N$,
$N=\dim \pi_\ell H$.  The diffusion process is thus degenerate with the
dimension $D$ of the noise (much) smaller than $N$. The strong Feller
property follows for such diffusions provided the generator
of the diffusion process is hypoelliptic. Let us discuss this next.

Recall the Fourier basis $\{e_k\}$ for $H$. Let the range of
$\pi_\ell$ be the span of $\{e_k\}$ with $|k|\leq M$. Write
$\omega=\sum_k\omega_ke_k$. Then the equation
(\ref{gal}) reads
\qq
d\omega_k =v_k(\omega)dt+ \gamma_kd\beta_k.
\label{gal1}
\qqq
where $v_k$ is  given by%quadratic (plus a linear term from the diffusion) in $\omega$.
 \qq
v_k(\omega)= - \nu |k|^2 \omega_k - {1\over 8\pi^2} \! \sum_{j + \ell = k} (j_1\ell_2-j_2\ell_1) \Bigl({1\over |\ell|^{2}} - {1\over |j|^{2}}\Bigr)
w_j w_\ell
%\label{detail}
\non
\qqq
and $w_k=\hf\omega_{-k}+\frac{1}{2i}\omega_k$ for $k\in Z^+$ and
$w_{-k}=\bar w_k$.
The generator of this diffusion is 
\qq
 L=X_0+\sum_{k\in K}X_k^2
\label{L}
\qqq
where we recall that $\gamma_k=0$ for $k\notin K$. The vector fields $X_\alpha$
are given by
\qq
 X_0&=&\sum_kv_k\partial_{\omega_k}\non\\
 X_k&=&\gamma_k\partial_{\omega_k}\non
%\label{}
\qqq
An  operator of the form (\ref{L}) with smooth vector fields $X_\alpha$
is known to generate a semigroup $\CP_t$ with
 smooth kernel (hence it is strong Feller) provided
 the  H\"ormander bracket
condition is satisfied ($L$ is then hypoelliptic). The condition is that
the span of the vector fields $X_j$, $j\neq 0$ and $[X_{i_1},[X_{i_2},\dots [X_{i_{k-1}},X_{i_k}]]\dots] $
for $k>1$ and $i_j\in\{0\}\cup K$ at each $\omega\in\R^N$ equals $R^N$.

To check this condition in the NS case  is a purely algebraic excercise
and the result is
the following \cite{EM,HM1}: 

\prop The following conditions for the set
$K\subset \IZ^2\setminus \{0\}$ are sufficient for  the H\"ormander bracket
condition to be satisfied:

(a) $K$ is invariant under the reflection $k\to -k$

(b) $K$ contains at least two elements of unequal length 

(c) $K$ spans $\IZ^2$ under linear combinations with integer coefficients.

\eprop

An example of a very degenerate forcing that suffices is given by the 
the set $K=\{(1,0),(-1,0), (1,1), (-1,-1)\}$ i.e there is forcing only on
two wave vectors and their reflections.

Note that the Proposition 5.1. is true for arbitrary (large enough)  Galerkin cutoff $N$.
Hence the full infinite dimensional generator formally satisfies the
H\"ormander condition and one might be tempted to try  to use this
 to return to the attempt to prove the strong  Feller property for $\CP_t$.
 However, it is likely that, as we let $N$ increase, the derivatives
 of the kernel of $\CP_t$ with respect to the high modes blow up
 since the smoothing is very weak for them. It is much more
 natural to try to use in that regime the dissipation as coded in
 the asymptotic strong Feller condition.
 
Let us finally remark that if all the  $\gamma_k$ in (\ref{gal1})
are 
nonzero the generator $L$ is elliptic. If $N$ is large enough
(of the order $\epsilon/ \nu^3$)  then
one may use the dissipativity of the high mode dynamics
to solve for the high modes in terms of the (temporal history)  of the
low modes and use the ellipticity of the latter to prove ergodicity and
mixing of the full dynamics \cite{BKL,EMS,KS}. 

\section{MALLIAVIN MATRIX}

Why does elliptic diffusion produce smoothness in transition kernels?
One way to think about this is to consider trajectories of the flow. Noise
will make the trajectories non-unique: a change in the initial condition
can be compensated by the noise. In elliptic diffusions  noise
spans the whole space and the compensation is immediate, in hypoelliptic
diffusions  the nonlinearity spreads the noise
in all directions thanks to the bracket condition. Thus a derivative
of the solution in the initial condition should equal its derivative
in a particular direction in the (history of) noise space. Since we are
integrating over the noise the latter derivative can be integrated by parts
and hence an estimate like the strong Feller property can emerge.

To be more explicit  recall that we wrote the solution of the stochastic NS
equation as $\omega(t)=\Phi_t(\beta,\omega_0)$ with  $\Phi_t$
smooth in the noise $\beta\in C([0,\infty)), \R^D)$ and the  initial condition $\omega_0$.
Also we have denoted the derivative in the initial condition by $\langle D_{\omega_0}\omega(t),
\xi\rangle=J_{0,t}\xi=\xi(t)$ for $\xi\in H$. Thus
\qq
  \langle D \CP_t\phi(\omega_0),\xi\rangle=
  \E\langle (D\phi)(\omega(t)),\xi(t)\rangle.
\label{deri}
\qqq
Consider next the  infinitesimal change in  
the solution corresponding to the change  of the noise $\beta$ in the direction
$V\in C([0,\infty)), \R^D)$: $ \langle D_\beta\omega(t),V\rangle:=\zeta(t)$.
$\zeta(t)$ satisfies the same linearized NS equation (\ref{JLinear})
 but with with forcing $QV$:
\begin{equation}\label{eq:JLinear1}
d\zeta(t) =\CL_{\omega(t)}\zeta(t)dt+QdV(t)
\end{equation}
and zero initial condition. The natural space to vary the noise is the Cameron-Martin space
i.e.
to take $V$ of the form $V(t)=\int_0^t v(s)\,ds$
 with  $v\in L^2_{\rm loc}([0,\infty],\R^D)$. 
 By variation of constants
$\zeta(t)$ is then given by
\qq
 \zeta(t)=\int_0^tJ_{s,t}Qv(s)ds:=A_{t}v.
\label{var}
\qqq
Actually, the $v$ one will eventually use is itself a function of the
noise (see eq. (\ref{vdef1}})), but it will be a.s. in $L^2_{loc}$.
The upshot is that    $A_t: L^2([0,t],\R^D )\to H$ is an a.s.  bounded random operator and
so  the Frechet
derivative can be written as $$ \langle D_\beta\omega(t),V\rangle =\sum_{k\in\CK}\int_0^t\CD^k_s\omega(t)v_k(s)ds$$ where  the operator $\CD_s^k$ is called the Malliavin derivative and heuristically
corresponds to an instantaneous kick at time $s$ to the
direction $k$ in noise space. Explicitly
\qq
\CD^k_s\omega(t)=J_{s,t}\gamma_ke_k.
\label{ds}
\qqq

  Suppose now we can find
 a $v$ such that
 \qq
\xi(t)= \zeta(t)\ \ {\rm i.e.} \ \ J_{0,t}\xi=A_{t}v
\label{mal}
\qqq
Inserting this to eq. (\ref{deri})  we get
\qq
  \langle D \CP_t\phi(\omega_0),\xi\rangle=
  \E\langle (D\phi)(\omega(t)),\zeta(t)\rangle
  =\E \langle D_\beta\phi(\omega(t)),V\rangle.
\label{deri1}
\qqq
The derivative $D_\beta$ in eq.  (\ref{deri1}) can be integrated by parts in the
Gaussian Wiener measure  to obtain 
\qq
\E \langle D_\beta\phi(\omega(t)),V\rangle=\E( \phi(\omega(t)) D^*_\beta V).
\label{deri2}
\qqq
In other words, the expression $ D^*_\beta $ is the adjoint of $D_\beta$
in $L^2(\Omega,W)$.  If the process $v$ is adapted to the Brownian filtration its expression is
simply
$D^*_\beta V=\sum_k\int v_k(s)d\beta_k(s)$, the Ito integral. Otherwise a derivative of
$v$ with respect to the noise also appears and $D^*_\beta V$ is called  the Skorokhod integral 
of $v$. Combining (\ref{deri2}) with  (\ref{deri1}) 
the desired bound follows:
\qq
  |\langle D \CP_t\phi(\omega_0),\xi\rangle|
  \leq \|\phi\|_\infty \E |D^*_\beta V |.
\label{deri3}
\qqq
It remains to solve eq.  (\ref{mal}) for $V$ (i.e. for $v$). Let $A_t^*$ be
the  Hilbert
space adjoint of  $A_t$ i.e. explicitly 
\qq
(A_t^*\xi)(s)=Q^*J^*_{s,t}\xi
\label{astar}
\qqq
for $s\leq t$. Then
the Malliavin matrix is defined by 
 \qq 
 M(t):=A_tA_t^*=\int_0^tJ_{s,t}QQ^*J_{s,t}^*ds
\label{mt}
\qqq 
Suppose $M(t)$ is invertible. Then, clearly a solution to (\ref{mal}) is
given by  
\qq
v=A_{t}^* M_{0,t}^{-1}J_{0,t}\xi
\label{vdef}
\qqq

To sketch the rest of the story in the finite dimensional setup we need a
 bound for the Skorokhod integral appearing in eq. (\ref{deri3}) \cite{Nua}:
 \qq
\E (D^*_\beta V)^2\leq\E\int_0^t|v(s)|^2ds+\sum_{kl}\E\int \CD^k_sv_l(r)\CD^l_rv_k(s))dsdr.
\label{skob}
\qqq
 The first term is the usual identity for the $L^2$ norm of the Ito integral, the second term
 appears for a non-adapted $v$, as is the one given by (\ref{vdef}). To compute the
 Malliavin derivative of $v$ in (\ref{skob}) note that all we need is to compute $\CD_rJ_{s,t}$
 since $A_t$ and $M(t)$ are expressed in terms of  $J_{s,t}$. This in turn is obtained
 by differentiating the equation (\ref{JLinear}): $\eta:=\CD^k_r\xi$ satisfies
 \begin{equation}
\partial_t \eta(t) = \nu \Delta \xi(t)+ \CA\omega(t)\cdot\nabla\xi(t)+
\CA\xi_t\cdot\nabla\omega(t):=\CL_{\omega(t)}\eta(t)+B(J_{r,t}\gamma_ke_k,\eta(t))
\non%\label{JLinear5}
\end{equation}
where $B$ is the bilinear form in appearing in $\CL$. By variation of constants an
expression involving only $J$ emerges.
%Using the semigroup property $J_{0,t}=J_{s,t}J_{0,s}$ these relations can
%also be written as $M(t)=J_{0,t}\hat M(t)J_{0,t}^*$ with
%\qq 
 %\hat M(t):=\int_0^tJ_{0,s}^{-1}QQ^*{J_{0,s}^*}^{-1}
%\label{hatmt}
%\qqq 
%the so called reduced Malliavin matrix and $v(s)=Q^*{J_{0,s}^*}^{-1} \hat M(t)^{-1}\xi$. 

%In a finite dimensional setup (so e.g. for the Galerkin NS) $J_{0,t}$ is invertible and both formuli can be used. 
Thus in the finite dimensional setup (so e.g. for the Galerkin NS) the main work
to be done is to show that $M(t)^{-1}$ %(or $\hat M(t)^{-1}$) 
has good probabilistic bounds. Indeed it turns out $\|M(t)^{-1}\|$
is in $L^p(\Omega)$ for all $p<\infty$. In the infinite dimensional case with degenerate noise 
%In the infinite dimensional case with degenerate noise $J_{0,t}$
%is not invertible due to dissipation of high modes so we need
%to use the first definitions. 
it is unlikely that $M(t)$
is a.s. invertible. $QQ^*$ is proportional to the projection
in $H$ to the subspace generated by the noise. In the
expression for  $M(t)$ the dynamics  spreads the range beyond
this subspace, however we expect the projection of
the result to the high modes to be very small. The key estimate on the Malliavin matrix
Hairer and Mattingly prove is that $M(t)$ is unlikely to be small
on vectors that have large projection to low modes:
\prop
  For every
  $\alpha,\eta,p$ and every orthogonal projection $\pi_\ell$ on a
  finite number of Fourier modes{\rm ,} there exists $C$ such that
\begin{equation} \label{e:lemMP}
  \mathbb{P} \bigl(\inf_{\|\pi_\ell \phi\|
\ge \alpha\|\phi\|}\frac{(M\phi,\phi)}{ \|\phi\|^2} < \epsilon\bigr) \le C
  \epsilon^p\exp \bigl(\eta \|\omega_0\|^2\bigr)\;,
\end{equation}   
holds for every $\epsilon \in (0,1)${\rm ,} and for every $\omega_0 \in H$.
\eprop
We will not discuss the details of the proof which is the technical
core of the paper \cite{HM2} (see also \cite{MP}). However, for experts we want
make the following comments. One major difficulty H\&M face is
that the integrand in the expression for the Malliavin matrix
is not adapted, i.e. depends on the future noise. The usual way out
of this problem in the finite dimensional theory is to 
use the semigroup property $J_{0,t}=J_{s,t}J_{0,s}$ to rewrite $M(t)=J_{0,t}\hat M(t)J_{0,t}^*$ with
\qq 
 \hat M(t):=\int_0^tJ_{0,s}^{-1}QQ^*{J_{0,s}^*}^{-1}
\non%\label{hatmt}
\qqq 
the reduced Malliavin matrix (and the control $v(s)=Q^*{J_{0,s}^*}^{-1} \hat M(t)^{-1}\xi$). 
In finite dimensions $J_{0,t}$
is  invertible and now the integrand is adapted. The proof then uses
Norris lemma \cite{Nor} which states that if a semimartingale
is small then both its bounded variation part and local martingale part are small.
In the infinite dimensional case with degenerate noise $J_{0,t}$
is not invertible due to dissipation of the high modes. Hence one needs
to work with non-adapted processes. The way out for H\&M is the
polynomial nature of the nonlinearity. In the iterative proof to show
that $(\phi,M(t)\phi)$ small implies $s\to (J_{s,t}P(u(s))\phi,\phi)$ is small
for the various multiple commutators $P$ the $P$ will always be
a polynomial.  One then writes $u(s)=v(s)+Q\beta(s)$ where $v$ is more
regular and expands $P(u(s))$ in powers of $Q\beta(s)$, ending
up with a polynomial in the Wiener process $\beta(s)$ with coefficients
that are nonadapted processes, but with higher regularity. The
basic lemma one now needs is that such a Wiener polynomial
can be small only if all the coefficients are small (up to
events of small probability).

\section{LOW MODE CONTROL} 

The approach to prove smoothness sketched in
the previous section is a form of {\it stochastic control} where the noise is used
to force solution to a prescribed region in phase space (for results on stochastic
control in our setup see also \cite{AS}, \cite{AKSS}). 
We saw that an exact compensation of the change of initial condition by
a change in the noise seems impossible, but Proposition 6.1.  gives
reason to hope that partial compensation is possible for the low modes. 
Since by Lemma 5.2.  the high modes are contracted the idea of
Hairer and Mattingly is to do an approximate control such that 
instead of the full control (\ref{mal}) we have $\xi(t)- \zeta(t)\to 0$ as
$t\to\infty$. Thus, as before let $v\in L^2_{\rm loc}(\R_+,\R^D)$ be
a shift in the noise and $\zeta(t)=A_{0,t}v$ be the corresponding
(infinitesimal) shift in the solution. Let
\qq
\rho(t)=\xi(t)- \zeta(t).
\label{malapp}
\qqq
Then, instead of the identities (\ref{deri1}) and (\ref{deri2}) we obtain
\qq\label{deri4}
  \langle D \CP_t\phi(\omega_0),\xi\rangle&=&\E( \phi(\omega(t)) D^*_\beta V)+\E\langle D \phi(\omega(t)),\rho(t)\rangle\\
  &\leq&  \|\phi\|_\infty \E |D^*_\beta V |+ \|D\phi\|_\infty\E\|\rho(t)\|.\non
\qqq
The asymptotic strong Feller property will follow provided $v$ can be 
chosen such that $\E |D^*_\beta V |$ stays bounded  as $t\to\infty$ and
$\E\|\rho(t)\|$ tends to zero exponentially.

To find $v$ H\&M use use a construction where at successive time
intervals two steps are alternated, one where high modes contract, the
second where low modes are controlled by the noise. Suppose at some time $t$ we knew
$\rho(t)$ is mostly in the high mode subspace, i.e. $\|\pi_\ell\rho(t)\|<<\|\rho(t)\|$.
Then, at least for a short time it pays to set $v=0$ since the linearized dynamics
contracts such a $\rho$ strongly. However, we cannot do this for too
long since the low mode part of $\rho$ will increase. Then provided we can find a
$v$ that will compensate the low mode part on a fixed time interval while
leaving the high mode part approximately intact we can iterate the
procedure.

The low mode control is a simple modification of the full control explained in the previous section.  Let us take the time intervals as
$[n, n+1 ]$ with $n$ odd integer 
for the first step and even integer for the second step. Thus we set $v(t)=0$ for $t\in [n, n+1]$,
$n$ odd. Let $A_n:=A_{n, n+1}$, $M_n:=A_nA_n^*$ and
$J_n:=J_{n,n+1}$. For  $n$ even  take
\qq
v_n:=v|_{ [n, n+1]}=A_n^*(M_n+\beta)^{-1}J_n\rho(n).
\label{vdef1}
\qqq
Note that except for the parameter $\beta$ this agrees with the full control
(\ref{vdef}).  While for $\beta=0$ the inverse in (\ref{vdef1}) most likely 
doesn't exist, for $\beta>0$ it does. The point now is that for small enough
$\beta$  (\ref{vdef1}) does a good job for the low mode
control while the high modes remain approximately intact. To see this, compute
 \qq\label{lowc}
\rho(n+1)&=&\xi(n+1)-\zeta(n+1)\\
&=&J_n\xi(n)-(J_n\zeta(n)+A_nA_n^*(M_n+\beta)^{-1}J_n\rho(n))\non\\
&=&\beta(M_n+\beta)^{-1}J_n\rho(n)
\non
\qqq
By Proposition 6.1.
eigenvectors of $M_n$ with small eigenvalues have small projections
to the low modes. Hence one expects that for small $\beta$ the
operator $\beta(M_n+\beta)^{-1}$ is small on vectors $\psi$ with
$\|\pi_\ell \psi\|\geq \alpha\| \psi\|$ whereas it is obviously bounded by one
elsewhere. Combining the two steps we get the iteration
\qq
\rho(n+2)=J_{n+1}\beta(M_n+\beta)^{-1}J_n\rho(n)%:=\CM_n\rho(n).
\label{iter}
\qqq
Combining Lemma 5.2, the bound (\ref{JLinear2})  and Proposition 6.1. H\&M prove (\cite{HM1}, Lemma 4.16)
\prop
For every two constants $\gamma, \eta > 0$ and every $p\ge 1${\rm ,} there
  exists a constant $\beta_0 > 0$ such that for $n$ even
$$
  \E\bigl(\|\rho_{n+2}\|^p\,|\, \CF_n\bigr) \leq \gamma e^{\eta
    \|\omega_n\|^2} \|\rho_n\|^p
$$  
holds almost surely whenever $\beta \le \beta_0$.
\eprop
Iterating Proposition 8.1. the exponential decay of $\E\|\rho(t)\|$
then follows   (\cite{HM1}, Lemma 4.13).

What remains is to bound the term $ \E |D^*_\beta QV |$ in (\ref{deri4})
uniformly in $t$, i.e. to bound the two integrals in (\ref{skob}). The
crux of the matter here is that both terms can be written
as a sum over $n$ of factors proportional to $\rho(n)$ which
provides a convergence factor. For the first term this is obvious
by (\ref{vdef1}). For the second one we need to go back to
the integration by parts formula eq. (\ref{deri2}). By construction
$V(t)=\int_0^tv(s)ds=\sum_nV_n$ where $V_n$ is $\CF_{n+2}$
measurable. Thus since the integration by parts is local in time
$D^*_\beta V=\sum_nD^*_{\beta|_{[n,n+2]} }V_n$ and the
second factor becomes
\qq
\sum_n\E\int_ {[n,n+2]^2}\tr(\CD_sv(r),\CD_rv(s))dsdr.
\label{skob1}
\qqq
For details of how to finish the argument we refer the reader to Section 4.8. in \cite{HM1}.

\section{TURBULENCE}
We have seen that the NS dynamics has a unique stationary state under
very general conditions of the forcing. Moreover, it can be proven that
the dynamics is mixing \cite{HM3} and the stationary state is reached
exponentially fast from arbitrary initial conditions and for arbitrary large
Reynolds numbers $R$ (for earlier proofs of mixing
in the case where $R$-dependent number of modes are forced
see \cite{BKL}, \cite{Mat}). Does this mean we have reached
some understanding on the properties of this state, in particular on the phenomenon
of turbulence? The proof outlined in the previous sections uses properties
of the system that have counterparts in the phenomenological theory
of turbulence. These are the dissipation of the high Fourier modes
and the transfer of the noise from the forced modes to the unforced ones due
to nonlinearity. The latter point is significant 
because most results of the NS dynamics are based on the energy and enstrophy conservation laws alone, and those bounds
would hold even if the nonlinearity was zero. Therefore, the properties of the latter are not used.

This being said it must be stressed that we have gained
very little understanding of the actual nature of the invariant state.
Crucial part of the proof is irreducibility which is based on the
fact that $\omega=0$ belongs to the support of every invariant measure.
Recall that this holds, because there is a small probability that the 
random forces are close to zero for any given time interval so the fluid
flow slows down due to viscosity. This is clearly not the true reason
one sees fast approach to stationarity in physical experiments. The mixing
times resulting from visits to the origin will be much larger than the
ones observed. To understand the real mechanism for mixing one
has to understand much better the transfer of energy and enstrophy
from the forcing scale to other scales. 

It was Kraichnan's observation  \cite{Kra} that we should expect this transfer
to be in two dimensions quite different from the three dimensional case.
In three dimensions, according to the Richardson-Kolmogorov picture
the forcing in low modes injects into the system energy  which is
transported due to the nonlinearity in NS equation to the higher
modes and eventually dissipated by the viscous term 
by large enough modes. %(in our case at scales of the order $|k|\sim k_d:=\nu^{3/4}\epsilon^{-1/4}$).
This transport of energy through scales in wave number space (i.e. $|k|:=\kappa$)
is called the Richardson energy cascade. In fact 
the theory predicts a constant flux of energy from the injection scale (in
our case 1) to the dissipation scale $ \kappa_\nu$ (these claims can be
formulated in terms of various correlation functions in the putative
stationary state, see e.g. the review  \cite{Kup}).
Kraichnan noted  that the existence in 2d of the second conserved quantity
of the inviscid flow, the enstrophy, means that one has to pose the question
at what scales (if any) energy and enstrophy are dissipated and if there
exist separate fluxes for the two. His observation was that the
fluxes of energy and enstrophy are to opposite directions,
energy flows towards {\it low} modes and enstrophy towards
high ones. Moreover, energy tends to be not dissipated at all
whereas enstrophy is dissipated at high modes like energy in
the 3d case. The presence of the two cascades, the {\it direct cascade}
of enstrophy and the {\it inverse cascade} of energy is very
well established both numerically 
\cite{Bof} and experimentally \cite{RDCE}. In what follows
we will point out a couple of mathematical questions regarding this picture
which would be nice to understand.

To state the Kraichnan picture more precisely
it is convenient to work on a torus of size $N$ i.e. $\mathbb{T}^2_N:=(\R/(2\pi N\IZ)^2$
rather than $N=1$ we had before. Of course by simple scaling
we can get rid of the $N$ at the expense of changing $\nu$ and
the forcing scale, but since the theory involves large separations
of the scales of dissipation, forcing and injection it is natural to
take $N$ large (eventually
to infinity)
we rather not do that. 
Consider now the NS dynamics on $\mathbb{T}^2_N$ with the random forcing
on Fourier modes of size $|k|\sim \kappa_f>>N^{-1}$ (observe that now $k\in (N^{-1}\IZ)^2$
i.e. $|k|\geq
1/N$). We shall add to the NS equation (\ref{o})
an extra  term that damps the low Fourier modes  more strongly than the viscous 
term does (note that $\nu k^2$ can be as small
as $\nu/N^2$). This is the Ekman friction term $-\tau\omega$ for $\tau>0$.
Stationary states for this system exist for the same reasons as before and 
uniqueness should follow in the presence of the friction as without provided the conditions of 
Proposition 5.1. hold. The Kraichnan theory
makes predictions on this stationary state, call it $\mu_{\nu,\tau,N}$ in the various limits $N\to\infty$,
$\tau\to 0$ and $\nu\to 0$. 

The starting point is conservation laws of energy and enstrophy
following from the  enstrophy balance equation (\ref{enspro}) and
a corresponding one for energy and taking into account the
extra friction term in the equation. Since the unique stationary
state is translation invariant these become local identities, for enstrophy
 \qq
 \nu\E_{\nu,\tau,N}(\nabla\omega(x))^2+\tau\E_{\nu,\tau,N}(\omega(x))^2=\epsilon
\label{ebal}
\qqq
and analogously for energy 
 \qq
 \nu\E_{\nu,\tau,N}(\nabla u(x))^2+\tau\E_{\nu,\tau,N}(u(x))^2=\epsilon'
\label{ebal1}
\qqq
with $\epsilon'$ the energy injection rate (per unit volume) which is
proportional to $\epsilon\kappa_f^{-2}$.  $\E_{\nu,\tau,N}$
denotes expectation in the measure $\mu_{\nu,\tau,N}$.

The first question to pose is what happens to the {\it viscous} dissipation
of energy and enstrophy as $\nu\to 0$. All the evidence points to
vanishing of energy dissipation
\qq
\lim_{\nu\to 0}\nu\E_{\nu,\tau,N}(\nabla u(x))^2=0
\label{eds}
\qqq
Enstrophy dissipation is more subtle as we will see below,
but again it is believed \cite{Ber}  that it vanishes:
\qq
\lim_{\nu\to 0}\nu\E_{\nu,\tau,N}(\nabla \omega(x))^2=0
\label{endis}
\qqq
It would be interesting to prove these statements and also to understand
whether a limiting measure $\lim_{\nu\to 0}\mu_{\nu,\tau,N}$ exists
and is supported on solutions of the damped randomly forced  Euler equation. 
Indeed, some indications that this could be done comes from the work \cite{CR}
where time averages of solutions and statistical solutions are controlled in that limit.
They are shown to be given in terms of solutions of the Euler
equation and in particular  \cite{CR} prove the relation (\ref{endis}) in that setup.

The main predictions of the Kraichnan theory come from the 
limit  $N\to\infty$ and
$\tau\to 0$.
The limit $N\to\infty$ means we are considering the NS
dynamics in $\R^2$. It is an interesting problem to try to prove that
the (weak) limit $\lim_{N\to\infty}\mu_{\nu,\tau,N}=\mu_{\nu,\tau,\infty}$ exists.
Note that we don't expect this state to be supported on $L^2$ but rather on
polynomially bounded (and presumably smooth) functions. The reason the
large volume limit might exist is the damping of the low modes by the friction term.
It produces an effective low wave number cutoff (which turns out to be
$\sim\tau^{3/2}{\epsilon'}^{-\hf}$).

Granting this,
what happens if we now take $\tau\to 0$? 
Is there also a measure $\mu_{\nu,0,\infty}$? The prediction of the Kraichnan
theory is that the viscous energy dissipation (\ref{eds}) vanishes as 
$\nu\to 0$ {\it uniformly} in $\tau$. Thus in that limit $\E_{\nu,\tau,\infty}(u(x))^2=(\epsilon'-o(\nu))/\tau$
i.e. the average energy density is not bounded in the putative limiting measure $\mu_{\nu,0,\infty}$. However, it is believed that $\mu_{\nu,0,\infty}$ is supported on smooth
$\omega$ and in particular
$\lim_{\tau\to 0}\tau\E_{\nu,\tau,\infty}(\omega(x))^2=0$. Then (\ref{ebal}) implies
{\it dissipative anomaly} for enstrophy: enstrophy dissipation
remains nonzero as $\nu\to 0$ i.e. $\lim_{\nu\to 0}\nu\E_{\nu,0,\infty}(\nabla\omega(x))^2=\epsilon>0$.

The Kraichnan theory makes more quantitative predictions of the
distribution of energy and enstrophy according to wave number.  Define the {\it energy spectrum} for $\kappa\in \R_+$
\qq
e(\kappa)=2\pi \hat g(\kappa)/\kappa
\label{es}
\qqq
where $\hat g(|k|)$ is the Fourier transform of the vorticity 2-point function
$$g(x-y)=\E_{\nu,\tau,\infty}\omega(x)\omega(y).$$ Then
energy density is given by
\qq
\E_{\nu,\tau,\infty}u(x)^2=\int_0^\infty e(\kappa)d\kappa
\label{ed}
\qqq
and enstrophy density  by $
\int_0^\infty e(\kappa)\kappa^2d\kappa$. Kraichnan theory predicts
\qq\label{spect}
e(\kappa) \sim \begin{cases}%\varepsilon_{\omega}^{2/3}\eta^3 f(\eta \kappa),&
%\kappa >> \eta^{-1}\\
\varepsilon^{2/3}\kappa^{-3}, &\kappa_f << \kappa << \kappa_\nu\\
{\epsilon'}^{2/3}\kappa^{-5/3}, &\kappa_\tau<<\kappa << \kappa_f.
\end{cases}
\qqq
where $\kappa_\nu\sim\nu^{-\hf}\epsilon^{1\over 6}$ is the { dissipation scale}  and $\kappa_\tau\sim\tau^{3/2}{\epsilon'}^{-\hf}$ the friction scale. 

The picture painted by the Kraichnan theory on 2d turbulence is thus quite
complex. With well separated scales of viscous dissipation, injection and 
friction energy flows from the injection scale towards small wave numbers
and is eventually dissipated by the friction. In the absence of friction
and in infinite volume energy flows to ever smaller wave numbers 
and energy density is not defined in the stationary state.
Enstrophy in turn flows to high wave numbers and is dissipated
there by the viscosity. Only in the state $\mu_{\nu,0,\infty}$ as $\nu\to 0$
one expects to have constant fluxes of energy and enstrophy,
for some exact calculations (subject to regularity assumptions) see \cite{Ber}.
One has to be careful with the order
of limits as is seen from the behavior of enstrophy dissipation.
Note in particular that the stationary state $\mu_{\nu,0,N}$ which we have been
discussing in the previous sections does not exhibit turbulence
in the sense of cascades of energy and enstrophy. Here energy will
reside in low modes, indeed, in experiments one often  sees 
the formation of a few large vortices in the flow. If $\nu$ is taken
to zero in this state then both energy and enstrophy will blow up and indeed, no
limit measure exists \cite{Kuk}. In \cite{Kuk} it is proven that only by taking
the injection rate $\epsilon$ (and thus also $\epsilon'$) proportional
to $\nu$ a nontrivial limiting measure exists. Formally this limit still 
corresponds to diverging Reynolds number, but one does not expect it
to be a turbulent state with near constant fluxes of energy and enstrophy.

What makes the Kraichnan theory intriguing is that e.g. the spectrum
(\ref{spect}) seems to be very well verified numerically and experimentally. Moreover,
the invariant measure seems to possess strong scale invariance properties,
at least in the inverse cascade regime. There are even indications of conformal
invariance \cite{BBCF}. Thus it is not excluded that some of its properties could be
mathematically accessible.

\end{document}